\documentclass[12pt]{article}
\usepackage{amsthm,amssymb,amsmath}
\usepackage[english]{babel}

 1 
1

\newcommand{\bB}{\mathcal{B}}
\newcommand{\bL}{\mathcal{L}}
\newcommand{\bM}{\mathcal{M}}

\renewcommand{\d}{\operatorname{d}}

\newcommand{\be}{\begin{equation}}
\newcommand{\ee}{\end{equation}}
\begin{document}
\title{\sc Solutions
of the Dispersionless Toda Hierarchy Constrained by String
Equations
\thanks{Partially supported by DGCYT
project BFM2002-01607 }}
\author{Luis Mart\'{\i}nez Alonso$^{1,\ddag}$
 and  Elena Medina$^{2,\maltese}$\\
\emph{ $^1$Departamento de F\'{\i}sica Te\'{o}rica II, Universidad
Complutense}\\ \emph{E28040 Madrid, Spain} \\
\emph{$^2$Departamento de Matem\'{a}ticas, Universidad de
C\'{a}diz}\\\emph{ E11510, Puerto Real, C\'{a}diz, Spain}\\
\texttt{$^\ddag$luism@fis.ucm.es}\\
\texttt{$^{\maltese}$elena.medina@uca.es}}
\date{} \maketitle
\begin{abstract}
Solutions of the Riemann-Hilbert problem implementing the
twistorial structure of the  dispersionless Toda (dToda)
hierarchy are obtained. Two types of string equations are considered
which characterize solutions arising in hodograph sectors and
integrable structures of two-dimensional quantum gravity
and Laplacian growth problems.

\end{abstract}

\vspace*{.5cm}

\begin{center}\begin{minipage}{12cm}

\end{minipage}
\end{center}
\newpage

\section{Introduction}

The dispersionless Toda (dToda) hierarchy  \cite{1}-\cite{3}
describes  several relevant integrable structures as  the
genus zero-limit of the Landau-Ginzburg
formulation of two-dimensional string theory \cite{4}-\cite{6},
the dynamics of conformal maps and the Laplacian growth problem governing interface dynamics
\cite{7}-\cite{8}. It can be formulated
in terms of two Laurent series

\begin{equation}\label{1.2}
\bL=p+\sum_{n\geq 0}\frac{u_{n+1}(t,\bar{t},s)}{p^n},\quad\quad
{\bar{\bL}}\;^{-1}=\frac{\bar{u}_0(t,\bar{t},s)}{p}+ \sum_{n\geq
0}\bar{u}_{n+1}(t,\bar{t},s)p^n,
\end{equation}
with coefficients depending on the variables
$t:=(x:=t_1,t_2,\ldots),\,
\bar{t}:=(y:=\bar{t}_1,\bar{t}_2,\ldots)$ and the spacial variable
$s$,  and such that the following Lax equations are satisfied
\begin{equation}\everymath{\displaystyle}\label{1.3}
\left\{
\begin{array}{cc}
\frac{\partial \bL}{\partial
t_n}=\{\mathcal{B}_n,\bL\},&
\frac{\partial \bar{\bL}}{\partial t_n}=\{\mathcal{B}_n,\bar{\bL}\},\\\\
\frac{\partial \bL}{\partial \bar{t}_n}=
\{\bar{\mathcal{B}}_n,\bL\},& \frac{\partial
\bar{\bL}}{\partial \bar{t}_n}=
\{\bar{\mathcal{B}}_n,\bar{\bL}\},
\end{array}
\right .
\end{equation}
where the Poisson bracket is defined as
\begin{equation}\label{1.5}
\{F,G\}:=p\;\Big(\frac{\partial F}{\partial p}
\frac{\partial G}{\partial s}
-\frac{\partial F}{\partial s}\frac{\partial G}{\partial p}\Big).
\end{equation}
and
\begin{equation}\label{1.4}
\mathcal{B}_n:=(\mathcal{L}^n)_{\geq 0},\quad\quad
\bar{\mathcal{B}}_n:=(\bar{\mathcal{L}}\;^{-n})_{\leq -1}.
\end{equation}
Here  $(\ldots)_{\geq 0}\;((\ldots)_{\leq -1})$ means the part of the
Laurent series with positive (strictly negative) powers of $p$

The main integrable model included in the dToda hierarchy is the
Boyer-Finley equation \cite{9}
\begin{equation}\label{1.1}
\frac{\partial^2 U}{\partial x\partial y}+\frac{\partial^2 }
{\partial s^2}\Big(\exp U\Big)=0,
\end{equation}
where $U=U(x,y,s):=\ln\bar{u}_0$. It is a much studied
(2+1)-dimensional integrable system which appears in the
classification of self-dual Einstein spaces with rotational
Killing symmetry \cite{9}-\cite{11} and in the twistor analysis of
Einstein-Weyl spaces \cite{12,13}.

Several methods for finding solutions of the members of the dToda
hierarchy have been proposed. A general strategy within the
framework of dispersionless integrable hirarchies is the hodograph
method \cite{14}-\cite{17}. Another approach, used for generating
solutions of the Boyer-Finley equation, is the group foliation
method of \cite{18,19}. In the present work we are concerned with
a third scheme: the twistor method of Takasaki-Takebe
\cite{1,20}. It  involves an extended Lax formalism with a pair
of Orlov functions $\bM$ and $ \overline{\bM}$ such that
\begin{equation}\label{1.6}
\{\bL,\bM\}=\bL,\quad\quad \{ \bar{\bL}, \overline{\bM}\}= \bar{\bL},
\end{equation}
with Laurent expansions
\begin{align}\label{1.7}
\nonumber \bM=&\sum_{n\geq 1}n\,t_n\,\bL^n+s+\sum_{n\geq 1}\frac{v_n(t,\bar{t},s)}{\bL^n},\\\\
\nonumber \overline{\bM}=&-\sum_{n\geq 1}n\,\bar{t}_n \bar{\bL}\;^{-n} +s+\sum_{n\geq
1} \bar{v}_n(t,\bar{t},s) \bar{\bL}^n,
\end{align}
and verifying the same Lax equations \eqref{1.3} as $\bL$ and
$\bar{\bL}$. Solutions of the dToda hierarchy are characterized by
imposing two constraints (\emph{string equations})
of the form
\begin{equation}\label{1.9}
 P(\bL,\bM)=\bar{P}(\bar{\bL},\overline{\bM}),\quad\quad
 X(\bL,\bM)=\bar{X}(\bar{\bL},\overline{\bM}),
\end{equation}
where $(P(p,s),X(p,s))$ and $(\bar{P}(p,s),\bar{X}(p,s))$
are pairs of canonically
conjugate variables (i.e. $\{P,X\}=P,\;  \{\bar{P},\bar{X}\}
=\bar{P}$), which together with the conditions \eqref{1.2} and \eqref{1.7}
constitute a Riemann-Hilbert problem.

 In the present work we solve two types of systems of string equations
and obtain solutions of the
\emph{truncated} $(n_1,n_2)$-dToda hierarchy with a finite number of
nonzero times
\begin{align*}
t_n&= \bar{t}_m= 0,\quad  n> n_1\;\;,m
> n_2,\\
t_{n_1}&\neq 0,\; \bar{t}_{n_2}\neq 0.
\end{align*}
The first type of systems is characterized by a string equation
of the form
\begin{equation}
\bar{P}(\bar{\bL})=P(\bL).
\end{equation}
The corresponding solutions  verify reductions of the dToda
hierarchy depending on a finite number of unknown functions
so that they represent  \emph{hodograph sectors}. In Section 3
we show a natural way of deriving hodograph
relations in the twistor method and, as an illustration, several
solutions of \eqref{1.1} corresponding to  cases of the form
\[
\bar{\bL}\;^{\pm \beta}=\bL^{\alpha},
\]
are exhibited.

 Section 4 is devoted to the family  of systems of string equations
\begin{equation}\label{1.10}
\bar{\bL}^{\beta}=\alpha\,\frac{\bL^{\alpha}}{\bM},\qquad
\frac{1}{\beta}\overline{\bM}=\frac{1}{\alpha}\bM,
\end{equation}
introduced  by Takasaki \cite{6} for describing the
integrable structure of two-dimensional quantum gravity. The case
$\alpha=\beta=1$ is specially important since
the corresponding solutions of the dToda hierarchy describe the
genus zero-limit of two-dimensional quantum gravity coupled to
$c=1$ matter \cite{6}. Furthermore, it also underlies the
integrable structures in the theory of conformal maps \cite{7} and
 Laplacian growth problems \cite{8}. We obtain solutions of the $(n_1,n_2)$-dToda hierarchy
satisfying \eqref{1.10} with
\[
n_1>\alpha-\beta, \quad n_2>\beta-\alpha.
\]
These solutions are determined by a system of implicit
equations which can be conveniently handed with computer aid.
For all the cases considered we provide an
equation of hodograph type characterizing the function $\bar{u}_0$, so that a
corresponding solution of the Boyer-Finley equation
is obtained.

\section{The twistor method of solution}

In terms of  the differential form
\begin{equation}\label{2.3}
\omega:=d \ln p\wedge d s+
\sum_{n\geq 1}d  \bB_n\wedge d t_n + \sum_{n\geq 1}
d   \bar{\bB}_n\wedge d \bar{t}_n.
\end{equation}
the dToda hierarchy can be formulated
in the following concise way
\begin{equation}\label{2.4}
 \omega=d\ln\bL\wedge d\bM=d\ln \bar{\bL}\wedge
d \overline{\bM}.
\end{equation}
From \eqref{2.3} and \eqref{2.4} it follows the existence of two
functions $S=S(\bL,t,\bar{t},s)$ and
$ \bar{S}= \bar{S}( \bar{\bL},t,\bar{t},s)$ such that
\begin{align}\label{2.6}
\nonumber
dS=\bM d\ln \bL+\ln p \,d s+\sum_{n\geq 1}\bB_n dt_n+
\sum_{n\geq 1}\bar{\bB}_n d\bar{t}_n,\\\\
\nonumber d \bar{S}= \overline{\bM} d\ln  \bar{\bL}+\ln p\, d s+
\sum_{n\geq 1}\bB_n dt_n+ \sum_{n\geq 1}\bar{\bB}_n d\bar{t}_n,
\end{align}
so that they can be assumed to admit expansions of the form
\begin{align}\label{2.7}
\nonumber S=&\sum_{n\geq 1} t_n\bL^n+s\ln \bL-\sum_{n\geq
1}\frac{v_n(t,\bar{t},s)}{n}\bL^{-n},\\\\
\nonumber  \bar{S}=&\sum_{n\geq 1}\bar{t}_n \bar{\bL}^{-n}+s\ln
 \bar{\bL}+\Phi(x,y,s)+\sum_{n\geq
1}\frac{ \bar{v}_n(t,\bar{t},s)}{n} \bar{\bL}^{n}.
\end{align}

The twistor method for solving the dToda hierarchy is supplied by
the following result \cite{3}

\vspace{0.3truecm}
\noindent
{\bf Theorem}\emph{ Let $(P(p,s),X(p,s))$ and $(\bar{P}(p,s),\bar{X}(p,s))$  be a pair of canonically
conjugate variables (i.e. $\{P,X\}=P,\;  \{\bar{P},\bar{X}\}
=\bar{P}$). If $(\bL,\bM,\bar{\bL}, \overline{\bM})$ are functions
of $(p,t,\bar{t},s)$ which admit expansions of the form \eqref{1.2}
and \eqref{1.7}, and satisfy the equations
\begin{equation}\label{2.8}
 P(\bL,\bM)=\bar{P}(\bar{\bL},\overline{\bM}),\quad\quad
 X(\bL,\bM)=\bar{X}(\bar{\bL},\overline{\bM}),
\end{equation}
then $(\bL,\bM,\bar{\bL}, \overline{\bM})$ is a solution of the dToda hierarchy.
}

\vspace{0.3truecm}

At this point two remarks are in order

\begin{enumerate}
\item The string equations \eqref{2.8} are meaningful only when they
are interpreted as a suitable Riemann-Hilbert problem on the complex
plane of the variable $p$. Indeed, $(\bL,\bM)$  must be analytic
functions in a neighborhood $D=\{|p|>r\}$ of $p=\infty$  and $(\bar{\bL},\overline{\bM})$ must be
analytic functions in a neighborhood $\bar{D}=\{|p|<\bar{r}\}$ of $p=0$. Thus the statement
of the Theorem  holds provided the string equations are satisfied on a common annulus
$A\subset D\bigcap\bar{D}$. The proof (see \cite{3}) consists
on differentiating \eqref{2.8}
with respect to $(p,t,\bar{t},s)\; (p\in A)$, then using the
Laurent series
in $D$ and $\bar{D}$ to obtain two expansions of the results in powers of $p$
on $A$ and, finally, identifying coefficients of both expansions.

In the cases considered  below we
impose conditions for
\[
P(\bL,\bM),\,\bar{P}(\bar{\bL},\overline{\bM}),\,
X(\bL,\bM),\,\bar{X}(\bar{\bL},\overline{\bM}),
\]
to be  analytic functions of $p$ on $A:=\mathbb{C}-\{0\}$. These
conditions play an essential role in our method as they
constitute the relations describing the hodograph
sectors represented by our first class of solutions and  are
part of the conditions required to characterize the solutions of
Takasaki string equations.

\item It is helpful to use canonical
generating functions \cite{20,21} to introduce  pairs of
conjugate variables . For example,  the condition
of canonicity for a pair $(P,X)$
\[
p\; d P\wedge d X= P\; dp\wedge d s,
\]
is ensured by defining $(P,X)$ through generating functions
$J_0=J_0(P,s_1)$ or $J_1=J_1(P,p)$ verifying
\begin{equation}\label{2.9}
d J_0=\frac{X}{P}\,d P+p\; d s_1,\quad d J_1=\frac{X}{P}\,d P-s_1d
p,
\end{equation}
where
\[
 s_1:=\frac{s}{p}.
\]
Equivalently
\begin{align}\label{2.9a}
\nonumber p&=\frac{\partial J_0}{\partial s_1},\quad
X=P\;\frac{\partial J_0}{\partial P},\\\\
\nonumber s&=-p\;\frac{\partial J_1}{\partial p},\quad
X=P\;\frac{\partial J_1} {\partial P}.
\end{align}
\end{enumerate}

\section{Hodograph Sectors}

\subsection{Hodograph relations in the twistor formalism}

From  generating functions of the form
\begin{equation}\label{3.1}
J_0(P,s_1)=f(P)\,s_1+g(P),\quad\quad \bar{J}_0(\bar{P},s_1)=
\bar{f}(\bar{P})\,s_1,
\end{equation}
we determine two pairs of conjugate variables $(P,X)$ and
$(\bar{P},\bar{X})$
given by
\begin{align*}
&P=P(p),\quad X=s\,\frac{\d \ln\,p}{\d \ln P}+\frac{\d g(P)}{\d \ln P},\\
&\bar{P}=\bar{P}(p),\quad \bar{X}=
s\,\frac{\d \ln\,p}{\d \ln \bar{P}},
\end{align*}
where $P=P(p)$ and $\bar{P}=\bar{P}(p)$ are the inverse functions of $f=f(P)$
$\bar{f}=\bar{f}(P)$, respectively. It follows at once that the
corresponding  string equations  are
\begin{equation}\label{3.2}
 \bar{P}(\bar{\bL})=P(\bL),
\end{equation}
and
\begin{equation}\label{3.2a}
\overline{\bM}\frac{\partial\ln\bar{\bL}}{\partial \ln \bar{P}(\bar{\bL})}
=
\bM\frac{\partial\ln \bL}{\partial \ln P(\bL)}+
\frac{\partial g(P(\bL))}{\partial \ln P(\bL)}.
\end{equation}
The second string equation can be rewritten as
\begin{equation}\label{3.3}
 \overline{\bM}=\frac{\partial (S+g(\bar{P}(\bar{\bL})))}{\partial \ln \bar{\bL}},
\end{equation}
or, equivalently, in terms of derivatives with respect to the
variable $p$
\begin{equation}\label{3.4}
 \overline{\bM}=\frac{\partial_p (S+g(P(\bL)))}{\partial_p
P(\bL)}\,\bar{\bL}\,\,\bar{P}'(\bar{\bL}).
\end{equation}

 We may design a method of
solution of \eqref{3.2} and \eqref{3.4} provided the following conditions are satisfied :
\begin{description}
\item[A1)] There exists  a solution $\bL=\bL(p,w)$ and $ \bar{\bL}=
 \bar{\bL}(p,w)$ of \eqref{3.2}, of the form \eqref{1.2},
depending on a finite number $N$ of unknown coefficients $w:=(w_1,\ldots,w_N)$.
\item[A2)] The function $\partial_p\ln P(\bL(p,w))$ vanishes at exactly
$N$  different points  $p_i=p_i(w)$
\end{description}

Indeed, under these assumptions we determine the unknowns $w$ by
imposing
\begin{equation}\label{3.5}
S+g(P(\bL))=\sum_{n=1}^{n_1} t_n\,\bB_n+\sum_{n=1}^{n_2}
\bar{t}_n\,\bar{\bB}_n+s\ln p+ \Big(g(P(\bL))\Big)_{\geq 0},
\end{equation}
and the vanishing of the numerator of \eqref{3.4} at the points
$p_i$. Thus we get the $N$ \emph{hodograph} relations
\begin{equation}\label{3.6}
\sum_{n=1}^{n_1}t_n\,\partial_p\bB_n(p_i)+\sum_{n=1}^{n_2}
\bar{t}_n\,\partial_p\bar{\bB}_n(p_i)+\frac{s}{p_i} + h(p_i)=0,
\end{equation}
where
\[
h(p):=\partial_p \Big(g(P(\bL(p,w)))\Big)_{\geq 0},
\]
and find $ \overline{\bM}$ from \eqref{3.4}. Notice that the hodograph relations prevent $ \overline{\bM}$
from having poles at the points $p_i$. Moreover,  near $p=0$ the form of $ \overline{\bM}$ satisfies
\eqref{1.7} since from \eqref{3.5} and by taking into account the assumption {\bf A1)} we deduce
\[
S+g(P(\bL))=\sum_{n=1}^{n_2}
\bar{t}_n\bar{\bL}^{-n}+s\ln \bar{\bL}+\mathcal{O}(1)
,\quad \bar{\bL}\rightarrow 0,
\]
and therefore
\[
 \overline{\bM}=\frac{\partial (S+g(P(\bL)))}{\partial
\ln\bar{\bL}}=-\sum_{n=1}^{n_2}n\,
\bar{t}_n\bar{\bL}^{-n}+s+\mathcal{O}( \bar{\bL}).
\]
Similarly, the expression \eqref{3.5} for $S$ leads to a function $\bM$ with an expansion of the
form \eqref{1.7}.

We notice that for the simplest case $n_1=n_2=1$, \eqref{3.6}
becomes
\[
x+\frac{s}{p_i}-\frac{ \bar{u}_0(w)}{p_i^2}y+ h(p_i)=0,
\]
which coincides with the system of hodograph relations  for the Boyer-Finley
equation found in \cite{15}.

\subsection{Examples}

If we take
\begin{equation}\label{3.7}
 \bar{\bL}^{\beta}=\bL^{\alpha},\quad \alpha>\beta>0,
\end{equation}
then it is immediate to see that expressions of the form
\begin{align*}
 \bar{\bL}&=\Big(p^{\beta}\,w_1+\cdots+p^{\alpha-1}\,
w_{\alpha-\beta}
+p^{\alpha}\Big)^{\frac{1}{\beta}},\\
\bL&=\Big(p^{\beta}\,w_1+\cdots+p^{\alpha-1}\,w_{\alpha-\beta}
+p^{\alpha}\Big)^{\frac{1}{\alpha}}
\end{align*}
solve \eqref{3.7}, depend on the $\alpha-\beta$ unknown coefficients $w=(w_1,\ldots,w_{\alpha-\beta})$
and have expansions of the form \eqref{1.2}.
Furthermore,
\[
\partial_p\ln P(\bL)=\frac{\alpha p^{\alpha-\beta}+
(\alpha-1)p^{\alpha-\beta-1}\,w_{\alpha-\beta}+\cdots+\beta\, w_1}
{p^{\alpha-\beta+1}+\cdots +p\,w_1},
\]
has exactly $\alpha-\beta$ zeros $p_i=p_i(w)$. Hence, the
assumptions {\bf A1)} and {\bf A2)} are satisfied and the
hodograph relations \eqref{3.6} determine solutions of the
truncated dToda hierarchy.

For example if
\[
\alpha=2,\;\; \beta=1,\quad g(P)= 0,
\]
we have $w=w_1,\; \bar{u}_0=1/w_1$ and it follows an hodograph relation for
the $(1,1)$-dToda hierarchy of the form
\begin{equation}\label{3.8}
x-4\,y\, \bar{u}_0^3-2\,s\, \bar{u}_0=0,
\end{equation}
which provides the solution
\begin{align*}
\bar{u}_0=&\frac{s}{3^{\frac{1}{3}}}\,{\sqrt{y}}\,{\left( -9\,x\,{\sqrt{y}} +
{\sqrt{3}}\,{\sqrt{8\,s^3 + 27\,x^2\,y}}
         \right) }^{-\frac{1}{3}}\\ -&
         \frac{1}{2\cdot 3^{\frac{2}{3}}\,{\sqrt{y}}}
         \left( -9\,x\,{\sqrt{y}} +
        {\sqrt{3}}\,{\sqrt{8\,s^3 + 27\,x^2\,y}} \right)
        ^{\frac{1}{3}},
\end{align*}
Other solutions  of the $(1,1)$-dToda hierarchy are
\begin{description}
\item[1)] $\alpha=3,\quad \beta=1,\quad g(P)=0$
\[
w=(w_1,w_2),\quad \bar{u}_0=\frac{1}{w_1}.
\]
The system of hodograph equations for $w$ is
\begin{align*}
x-\frac{6\,s}{2\,{w_2} + {\sqrt{-12\,{w_1} + 4\,{{w_2}}^2}}}
-
  \frac{36\,y}{{w_1}\,{\left( -2\,{w_2} -
         {\sqrt{-12\,{w_1} + 4\,{{w_2}}^2}} \right) }^2}&=0\\\\
x-\frac{6\,s}{2\,{w_2} - {\sqrt{-12\,{w_1} + 4\,{{w_2}}^2}}}  -
  \frac{36\,y}{{w_1}\,{\left( -2\,{w_2} +
         {\sqrt{-12\,{w_1} + 4\,{{w_2}}^2}} \right) }^2}&=0
\end{align*}
which leads to the following implicit equation for $\bar{u}_0$
\begin{equation}\label{n01}
(x+3\,y\,\bar{u}_0^2)(-3\,s^2\,\bar{u}_0 + {\left( x -
3\,y\,\bar{u}_0^2 \right) }^2)=0.
\end{equation}
\item[2)] $\alpha=3,\quad \beta=2,\quad g(P)=0$
\[
w=w_1,\quad \bar{u}_0=\frac{1}{w_1^{1/2}}.
\]
We get the hodograph equation
\begin{equation}\label{n1}
 -6\,s\,\bar{u}_0^2 + 4\,x - 9\,\, y\,\bar{u}_0^5=0.\end{equation}

\item[3)] $\alpha=2,\quad \beta=1,\quad g(P)=P^{\frac{3}{2}}$
\[
w=w_1,\quad \bar{u}_0=\frac{1}{w_1}.
\]
The hodograph equation is
\begin{equation}\label{n2} -3 - 16\,s\,\bar{u}_0^3 +
8\,x\,\bar{u}_0^2 - 32\,y\,\bar{u}_0^5=0.\end{equation}
\end{description}

For the choice
\begin{equation}\label{3.9}
 \frac{1}{\bar{\bL}\;^{\beta}}=\bL^{\alpha}
,\quad \alpha,\beta>0,
\end{equation}
it follows that
\begin{align*}
\frac{1}{ \bar{\bL}}&=\Big(\frac{w_{-\beta}}{p^{\beta}}+
\cdots+\frac{w_{-1}}{p}+w_1+p\,w_2+\cdots+
p^{\alpha-1}\,w_{\alpha}+p^{\alpha}\Big)^{\frac{1}{\beta}},\\
\bL&=\Big(\frac{w_{-\beta}}{p^{\beta}}+
\cdots+\frac{w_{-1}}{p}+w_1+p\,w_2+\cdots+
p^{\alpha-1}\,w_{\alpha}+p^{\alpha}\Big)^{\frac{1}{\alpha}},
\end{align*}
solve \eqref{3.9}, depend on the $\alpha+\beta$ unknown
coefficients $w=(w_{-\beta},\ldots,w_{-1},w_1,\ldots,w_{\alpha})$
and have expansions of the form \eqref{1.2}.
Moreover,
\[
 \partial_p\ln P(\bL)=\frac{\alpha p^{\alpha+\beta}+
(\alpha-1)p^{\alpha+\beta-1}\,w_{\alpha}+\cdots-\beta\,
 w_{-\beta}}
{p^{\alpha+\beta+1}+\cdots+ p\,w_{-\beta}},
\]
has exactly $\alpha+\beta$ zeros $p_i=p_i(w)$. Hence, the
assumptions {\bf A1)} and {\bf A2)} are satisfied so that the
hodograph relations \eqref{3.6} determine solutions of the dToda
hierarchy.

For example if $\alpha=2,\quad\beta=1,\quad g(P)=P^{-3}$, then
$$
w=(w_{-1},w_1,w_2),\quad \bar{u}_0=w_{-1}.
$$
In this case the hodograph equations for $w$ lead to the system
$$\begin{array}{c}
12\,w_1\,w_{-1} + x=0,\\  \\
s + 6\,w_1\,w_2\,w_{-1} + 6\,w_{-1}^2=0,\\  \\
3\,w_1^2\,w_{-1} + 6\,w_2\,w_{-1}^2 -y\,w_{-1}=0,
\end{array}$$
which implies the following implicit relation for $\bar{u}_0$
\begin{equation}\label{n3}
- x^3 + 48\,x\,y\,\bar{u}_0^2 - 576\,s\,\bar{u}_0^3-3456\,\bar{u}_0^5 =0.
\end{equation}

\section {Takasaki String  Equations}

\subsection{General scheme}

By taking the generating functions
$$
J_1(P,p)=-\frac{p^{\alpha}}{P},\quad\quad \bar{J}_0(\bar{P},s_1)
=\bar{P}^{1/\beta}\,s_1,\quad \alpha,\beta>0,
$$
we determine the pairs of conjugate variables
\begin{align*}
&P=\alpha\,\frac{p^{\alpha}}{s},\quad X=\frac{s}{\alpha},\\
&\bar{P}=p^{\beta},\quad \bar{X}=\frac{s}{\beta}.
\end{align*}
They lead to the string equations proposed by Takasaki
\begin{equation}\label{tw0}
\bar{\bL}^{\beta}=\alpha\,\frac{\bL^{\alpha}}{\bM},\qquad
\frac{1}{\beta}\overline{\bM}=\frac{1}{\alpha}\bM,
\end{equation}
or equivalently
\begin{equation}\label{tw1}
\beta{\bM}=\alpha{\overline{\bM}},\end{equation}
\begin{equation}\label{tw2}
\beta\,\frac{\bL^{\alpha}}{\bar{\bL}\;^{\beta}}={\overline{\bM}}.\end{equation}
We next prove that the string equations \eqref{tw1}, \eqref{tw2} have solutions satisfying
\eqref{1.2} and \eqref{1.7} with
\begin{equation}\label{m1} \everymath{\displaystyle}
\begin{array}{rcl}
{\bM}&=&\sum_{n=1}^{n_1}nt_n\bL^n+s+\sum_{n=1}^{\infty}v_n\bL^{-n},\\  \\
{\overline{\bM}}&=&-\sum_{n=1}^{n_2}n\,\bar{t}_n \bar{\bL}\;^{-n}+s+\sum_{n=1}^{\infty}\bar{v}_n\bar{\bL}^{n}.
\end{array}
\end{equation}

Given two integers $r\leq s$ let us denote by
$\mathrm{V}[r,s]$ the set of truncated Laurent series of
the form
\[
c_r\,p^r+c_{r+1}\,p^{r+1}+\cdots+c_s\,p^s.
\]
Let us look for solutions of \eqref{tw1}-\eqref{tw2} such that
$\bL^{\alpha}$ and $\bar{\bL}^{-\beta}$ are meromorphic functions
of $p$ with possible poles only at $p=0$ and $p=\infty$. Then, as
a consequence of the assumptions \eqref{1.2},\eqref{m1} and  the twistor equations
\[
\frac{\alpha}{\bar{\bL}^{\beta}}=\bM\,\bL^{-\alpha},\quad
\beta\,\bL^{\alpha}=\overline{\bM}\,\bar{\bL}^{\beta},
\]
it follows that
\begin{equation}\label{con2}
\frac{1}{\bar{\bL}^{\beta}}\in\mathrm{V}[-\beta,n_1-\alpha],\quad
\bL^{\alpha}\in\mathrm{V}[\beta-n_2,\alpha].
\end{equation}
Thus, the existence of nontrivial solutions requieres
\begin{equation}\label{con}
n_1>\alpha-\beta, \quad n_2>\beta-\alpha.
\end{equation}

We can split  \eqref{tw1} into the
system of equations:
\begin{equation}\label{tw11}
\alpha\,{\overline{\bM}}_{\geq1}=\beta\,\bM_{\geq 1},
\end{equation}
\begin{equation}\label{tw12}
\alpha\,{\overline{\bM}}_0=\beta\,\bM_0,
\end{equation}
\begin{equation} \label{tw13}
\alpha\,{\overline{\bM}}_{\leq -1}=\beta\,{\bM}_{\leq-1},\end{equation}
where  $(\ldots)_{\geq 1}\;((\ldots)_{\leq -1})$ denote the part
of the Laurent series with strictly positive (strictly negative) powers of $p$, and
$(\ldots)_0$ stands for the constant term.
Obviously this system is satisfied if we set
\begin{align} \label{alphabm1}
\nonumber \bM=\sum_{n=1}^{n_1}nt_n(\bL^n)_{\geq1}+
\frac{\alpha}{\beta}\,\overline{\bM}_0
- \frac{\alpha}{\beta}\,\sum_{n=1}^{n_2}n\,\bar{t}_n
(\bar{\bL}\;^{-n})_{\leq -1},\\\\
\nonumber {\overline{\bM}}=-\sum_{n=1}^{n_2}n\,\bar{t}_n
(\bar{\bL}\;^{-n})_{\leq -1}+\overline{\bM}_0+
\frac{\beta}{\alpha}\;\sum_{n=1}^{n_1}nt_n
(\bL^n)_{\geq1},
\end{align}
where
\[
\overline{\bM}_0=s-\sum_{n=1}^{n_2}n\,\bar{t}_n
(\bar{\bL}\;^{-n})_0.
\]
Moreover, from \eqref{alphabm1} it can be easily seen  that
${\overline{\bM}}$
has the required expansion of the form \eqref{1.7}
provided $\bL$ and $\bar{\bL}$ satisfy \eqref{1.2}. On the other hand,
the expression \eqref{alphabm1} for $\bM$ has an expansion of
the form \eqref{m1} if the residue of $\bM\,\bL^{-1}$
corresponding to its Laurent expansion in powers of
$\bL$ verifies

\begin{equation}\label{m0}
Res(\frac{\bM}{\bL},\bL)=s.
\end{equation}
Hence the problem reduces  to finding
$\bL$ and $\bar{\bL}$ satisfying
\eqref{1.2}, \eqref{tw2} and \eqref{m0}.

In view of \eqref{1.2} and \eqref{con2} we look for  $\bar{\bL}\;^{-\beta}$ and
$\bL^{\alpha}$ in the form
\begin{align} \label{blbeta}
\nonumber &\bar{\bL}\;^{-\beta}=\frac{\bar{w}_0}{p^{\beta}}
+
 \frac{\bar{w}_1}{p^{\beta-1}}+\cdots+\frac{\bar{w}_{m}}{p^{\beta-m}},\quad
 m:=n_1-\alpha+\beta,\\\\
\nonumber &\bL^{\alpha}=p^{\alpha}+w_1\,p^{\alpha-1}+
\cdots+w_{\alpha-\beta+n_2}\,p^{\beta-n_2}.
\end{align}
Hence \eqref{tw2} reads
\begin{equation}\label{hod1}
\beta\,\bL^{\alpha}=\frac{p^{\beta}\,\overline{\bM}}{
\bar{w}_m\,p^m+\cdots+ \bar{w}_1\,p +\bar{w}_0},
\end{equation}
and in order to prevent  $\bL^{\alpha}$ from having poles
different from $p=0$ and $p=\infty$ we impose
\begin{equation}\label{hod2}
\overline{\bM}(p_i(\bar{w}))=0,
\end{equation}
where $p_i(\bar{w}),\,(\bar{w}:=(\bar{w}_1,\cdots,\bar{w}_m))$ denote the $m$ zeros of
\[
\bar{w}_m\,p^m+\cdots+ \bar{w}_1\,p+\bar{w}_0=0.
\]

In this way by using \eqref{blbeta} in
the expression
\eqref{alphabm1} for $\overline{\bM}$, the equation \eqref{hod1}
becomes dependent on the variables
\[
(p,s,t,\bar{t},\bar{w}_0,\ldots,\bar{w}_m,w_1,
\ldots,w_{\alpha-\beta+n_2}).
\]
 Thus, by identifying coefficients
of the powers $p^i,\, i=\beta-n_2,\ldots,\alpha$ we get $\alpha-
\beta+n_2+1$ equations  which together with the $m$ equations
\eqref{hod2} determine the  $\alpha-\beta+n_2+m+1$   unknowns variables $(\bar{w}_0,
\ldots,\bar{w}_m,w_1,
\ldots,w_{\alpha-\beta+n_2})$ as functions of $(s,t,\bar{t})$. However,
to complete our proof we must show that \eqref{m0} is satisfied too. To do
that let us take two circles $\gamma\; (|p|=r)$ and $\bar{\gamma}\;
(|p|=\bar{r})$ in the complex $p$-plane  and denote by $\Gamma$ and
$\bar{\Gamma}$ their images under the maps $\bL=\bL(p)$ and
$\bar{\bL}=\bar{\bL}(p)$, respectively. Then we have
\begin{align*}
&Res(\frac{\bM}{\bL},\bL)-Res(\frac{\overline{\bM}}{\bar{\bL}},\bar{\bL})\\\\
&=\frac{1}{2i\pi}\oint_{\Gamma} \frac{\bM}{\bL}\d \bL-
\frac{1}{2i\pi}\oint_{\bar{\Gamma}}\frac{\overline{\bM}}{\bar{\bL}}
\d \bar{\bL}\\\\
&=\frac{1}{2i\pi}\oint_{\gamma} \frac{\partial_p \bL^{\alpha}}
{\bar{\bL}^{\beta}}\,\d p+
\frac{1}{2i\pi}\oint_{\bar{\gamma}} \bL^{\alpha}\,
\partial_p (\bar{\bL}\;^{-\beta})\,\d p
\\\\
&=\frac{1}{2i\pi}\oint_{\gamma} \partial_p\Big(\frac{\bL^{\alpha}}{\bar{\bL}^{\beta}}\Big)\,\d p=0,
\end{align*}
where we have taken into account that the integrands are analytic functions
of $p$ in $\mathbb{C}-\{0\}$ and that $\gamma$ and $\bar{\gamma}$
are homotopic with respect to $\mathbb{C}-\{0\}$.  Therefore,
as we have already proved that $\overline{\bM}$ has an
expansion of the form \eqref{1.7}, we deduce
\[
Res(\frac{\bM}{\bL},\bL)=Res(\frac{\overline{\bM}}{\bar{\bL}},\bar{\bL})=s,
\]
so that \eqref{m0} follows.

\subsection{Examples}

We first illustrate our method  by considering
two cases with $\alpha=\beta$.

\subsection*{I) $\alpha=\beta$, $n_1=n_2=1$}

The starting point is to set
$$\bL^{\alpha}=p^{\alpha}+w_1p^{\alpha-1},$$
$$\bar{\bL}\;^{-\alpha}=\frac{\bar{w}_0}{p^{\alpha}}+
\frac{\bar{w}_1}{p^{\alpha-1}}.$$
The polynomial $p^{\alpha}\bar{\bL}^{-\alpha}$ has a unique zero at
$$p_1=-\frac{\bar{w}_0}{\bar{w}_1},$$
thus, (46) leads us to
\begin{equation}\label{ex1eq1}
s - x\,{\frac{\bar{w}_0}{\bar{w}_1}} +
   \left(1 -\frac{1}{\alpha} \right)\,y \,
       {\bar{w}_0}^{-1 + \frac{1}{\alpha}}\,\bar{w}_1
   = 0
\end{equation}
Now, by equating the powers of $p$ in (45) we obtain
\begin{equation}\label{ex1eq2eq3}\everymath{\displaystyle}\begin{array}{rcl}
p^{\alpha}: &  &
\alpha{\bar{w}_1} +x = 0
  \\  \\
p^{\alpha-1}: &  &
\alpha\,w_1 - {\frac{s}{\bar{w}_1}} +
   {\frac{x\,\bar{w}_0}{{{\bar{w}_1}^2}}} +
   {\frac{{y\,{\bar{w}_0}^{-1 + {\frac{1}{\alpha}}}}}{\alpha}} = 0
\end{array}\end{equation}
Finally, from \eqref{ex1eq1} and \eqref{ex1eq2eq3}  we
can eliminate $\bar{w}_1$ and $w_1$, and taking into account that
$\bar{w}_0=\bar{u}_0^{\alpha}$ we get $\bar{u}_0$ implicitly
defined by
\begin{equation}\label{ab1}
(1-\alpha)\,x\,y\,\bar{u}_0-\alpha^2\,s\,\bar{u}_0^{\alpha}+\alpha^3\,\bar{u}_0^{2\alpha}=0.
\end{equation}
\subsection*{II) $\alpha=\beta$, $n_1=2$, $n_2=1$}

We start with the expressions
$$\bL^{\alpha}=p^{\alpha}+w_1p^{\alpha-1},$$
$$\bar{\bL}\;^{-\alpha}=\frac{\bar{w}_0}{p^{\alpha}}+
\frac{\bar{w}_1}{p^{\alpha-1}}+\frac{\bar{w}_2}{p^{\alpha-2}},$$
Now the polynomial $p^{\alpha}\bar{\bL}\;^{-\alpha}$ has two zeros at the
points
$$p_1={\frac{-\bar{w}_1+ {\sqrt{{{\bar{w}_1}^2} -
         4\,\bar{w}_0\,\bar{w}_2}}}{2\,\bar{w}_2}},
         \quad
p_2={\frac{-\bar{w}_1 - {\sqrt{{{\bar{w}_1}^2} -
         4\,\bar{w}_0\,\bar{w}_2}}}{2\,\bar{w}_2}},$$
thus, (46) yields two equations which become equivalent to
\begin{equation}\label{ex5eq1}\everymath{\displaystyle}\begin{array}{l}
2\,\alpha\,t_2\,\bar{w}_0\,{{\bar{w}_1}^2} -
   2\,\alpha\,t_2\,{{\bar{w}_0}^2}\,\bar{w}_2 -
   4\,t_2\,\bar{w}_0\,w_1\,\bar{w}_1\,
    \bar{w}_2 + \alpha\,s\,\bar{w}_0\,{{\bar{w}_2}^2}\\  \\
     -   \alpha\,x\,\bar{w}_0\,\bar{w}_1\,\bar{w}_2 -
   y\,{{\bar{w}_0}^{{\frac{1}{\alpha}}}}\,\bar{w}_1\,{{\bar{w}_2}^2}
   = 0,\end{array}
\end{equation}
\begin{equation}\label{ex5eq2}
2\,\alpha\,t_2\,{{\bar{w}_0}^2}\,\bar{w}_1 -
   4\,t_2\,{{\bar{w}_0}^2}\,w_1\,\bar{w}_2 -
   \alpha\,x\,{{\bar{w}_0}^2}\,\bar{w}_2 -
   \alpha\,y\,{{\bar{w}_0}^{1 + {\frac{1}{\alpha}}}}\,{{\bar{w}_2}^2} =
   0.
   \end{equation}
Now, by identifying coefficients  of powers of $p$ in (45) we obtain
\begin{equation}\label{ex5eq3eq4}\everymath{\displaystyle}\begin{array}{rcl}
p^{\alpha}: &  &
-2\,t_2 + \alpha\,\bar{w}_2 = 0
  \\  \\
p^{\alpha-1}: &  &
2\,a\,t_2\,\bar{w}_1 -
   4\,t_2\,w_1\,\bar{w}_2 +
   {\alpha^2}\,w_1\,{{\bar{w}_2}^2} - \alpha\,x\,\bar{w}_2 = 0
\end{array}\end{equation}
Finally, from \eqref{ex5eq1}, \eqref{ex5eq2} and \eqref{ex5eq3eq4}  we
can eliminate $\bar{w}_1$, $\bar{w}_2$ and $w_1$, and taking into account that
$\bar{w}_0=\bar{u}_0^{\alpha}$ we get the following implicit
equation for $\bar{u}_0$
\begin{equation}\label{ab2}
\everymath{\displaystyle}\begin{array}{l}
 \alpha^4\,s\,\bar{u}_0^{2\alpha} -
\alpha^5\,\bar{u}_0^{3\alpha}
 +\alpha^3\,x\,y\,\bar{u}_0^{\alpha +1}+
 4\,t_2\,y^2\,\bar{u}_0^2- 6\,\alpha\,t_2\,y^2\,\bar{u}_0^2
\\  \\
+  \alpha^2\,y\,\bar{u}_0( -x\,\bar{u}_0^{\alpha}+ 2\,t_2
   \,y\,\bar{u}_0)  = 0.
\end{array}
\end{equation}

\vspace{1cm}

Next we quote the final implicit relation for $\bar{u}_0$ corresponding
to some examples of solutions for $\alpha\neq\beta$.

\begin{description}

\item[1)] $\alpha=4$, $\beta=2$, $n_1=3$, $n_2=1$.

$$ 2560\,{\bar{u}_0}^7-1536\,t_2\,{\bar{u}_0}^5  +
  64\,\left( 2t_2^2 + 3\,t_3\,x \right)\,{\bar{u}_0}^3
  -72\,t_3^2\,s\,\bar{u}_0 -
  27{t_3}^3\,y=0.$$

\item[2)] $\alpha=3$, $\beta=2$, $n_1=2$, $n_2=1$.

$$
3\,s\,t_2\,\bar{u}_0 + 27\,{{\bar{u}_0}^5} -
   9\,x\,{{\bar{u}_0}^3} + {{t_2}^2}\,y = 0.$$
\item[3)] $\alpha=5$, $\beta=3$, $n_1=3$, $n_2=1$

$$\begin{array}{c}
  -120\,{t_3}^2\,s\,{\bar{u}_0}^2 +
  100\,{t_2}^2 \,{\bar{u}_0}^5 -
  2000\,t_2\,{\bar{u}_0}^8 + 4375\,{\bar{u}_0}^{11} \\  \\
  +  300\,t_3\,x\,{\bar{u}_0}^5 - 48\,{t_3}^3\,y=0.
\end{array}$$
\item[4)] $\alpha=2$, $\beta=1$, $n_1=3$, $n_2=1$.

$$
-3\,t_3\,s^2 - 16\,{\bar{u}_0}^3 + 8\,x\,{\bar{u}_0}^2 -
   12\,t_3\,y\,{\bar{u}_0}^2 = 0.$$

\item[5)] $\alpha=3$, $\beta=1$, $n_1=4$, $n_2=1$

$$\begin{array}{c}
   - 64\,{t_4}^3\,s^3 +  288\,t_2\,{t_4}^2\,s^2\,\bar{u}_0 -
   1296\,{t_4}^2\,s^2\,{\bar{u}_0}^2
   -   864\,{t_2}^3\,{\bar{u}_0}^3 \\  \\
  +  1296\,t_2\,t_4\,s\,{\bar{u}_0}^3 +7776
  \,{t_2}^2\,{\bar{u}_0}^4
  -  4860\,t_4\,s\,{\bar{u}_0}^4 -
   21870\,t_2\,{\bar{u}_0}^5  \\  \\
   + 18225\,{\bar{u}_0}^6+972\,t_4\,x^2\,{\bar{u}_0}^3
    -   2592\,{t_4}^2\,x\,y\,{\bar{u}_0}^3 +
   1728\,{t_4}^3\,y^2\,{\bar{u}_0}^3 = 0.\end{array}$$
\end{description}

\subsection{Applications to integrable contour dynamics}

Let $z=z(p)$ be an invertible conformal map of the exterior of the
unit circle $|p|>1$ to the exterior of a simply connected domain
bounded by a simple analytic curve $\gamma$. At $p\rightarrow\infty$ it can
be expanded in the form
\begin{equation}
z(p)=r\,p+\sum_{n=0}^{\infty}\frac{r_n}{p^n},
\end{equation}
where the coefficient $r$ is real. By expressing the coefficients $(r,r_0,r_1,\ldots)$ as functions of the
harmonic moments $t=(t_0,t_1,\ldots)$ of the exterior of $\gamma$
it turns out \cite{7}-\cite{8} that
the corresponding function $z(p,t)$ determines a solution of the dToda hierarchy.
The relation between the dynamical objects involved in the two different
gauges of the dToda hierarchy used in \cite{1}-\cite{3} and
\cite{6}-\cite{7} is as follows
\begin{equation}\label{cd1}
z(p)=\bL(r p),\quad \bar{z}(\frac{1}{p})=
\bar{\bL}\;^{-1}\Big(r p\Big),
\quad r:=\sqrt{\bar{u}_0},
\end{equation}
where
\begin{equation}\label{cd11}
\bar{z}(\frac{1}{p})=\frac{r}{p}+\sum_{n=0}^{\infty}\frac{r^*_n}{p^n},\quad
\end{equation}
and
\[
t_0=s,\quad \bar{t}_n=-t^*_n,\quad n\geq 1 .
\]
The associated system of string equations is
\begin{equation}\label{cd4}
\bar{\bL}=\frac{\bL}{\bM},\qquad
\overline{\bM}=\bM,
\end{equation}
which implies
\begin{equation}\label{cd2}
\{z(p),\bar{z}(\frac{1}{p})\}=1.
\end{equation}
Furthermore, in view of \eqref{cd1}-\eqref{cd11}, the solution satisfies the reduction
condition
\begin{equation}\label{cd5}
\frac{1}{\bar{\bL}(r p)}=(\bL(r p))^*,\quad\mbox{for $|p|=1$}.
\end{equation}
It can be seen that the method developed in Subsection 4.1 is
compatible with \eqref{cd5} provided $r$ is real, so that it
can be applied to obtain these solutions by setting
\[
\alpha=\beta=1,\quad n_1=n_2.
\]
The following two examples illustrate the simplest cases.

\subsection*{I) $n_1=n_2=2$ (the ellipse)}

The polynomial $p\,\bar{\bL}^{-1}$ has two zeros at the
points
$$p_1={\frac{-\bar{w}_1+ {\sqrt{{{\bar{w}_1}^2} -
         4\,\bar{w}_0\,\bar{w}_2}}}{2\,\bar{w}_2}},
         \quad
p_2={\frac{-\bar{w}_1 - {\sqrt{{{\bar{w}_1}^2} -
         4\,\bar{w}_0\,\bar{w}_2}}}{2\,\bar{w}_2}}.$$
From \eqref{hod2} we get two equations which lead to
\begin{equation}\label{ex8eq1}\everymath{\displaystyle}
\begin{array}{l}
-2\,t_2\,{{\bar{w}_1}^3} +
   4\,t_2\,\bar{w}_0\,\bar{w}_1\,\bar{w}_2 +
   t_1\,{{\bar{w}_1}^2}\,\bar{w}_2 +
   4\,t_2\,w_1\,{{\bar{w}_1}^2}\,\bar{w}_2 -
   t_1\,\bar{w}_0\,{{\bar{w}_2}^2} -
   4\,t_2\,\bar{w}_0\,w_1\,{{\bar{w}_2}^2}\\  \\
    -  t_0\,\bar{w}_1\,{{\bar{w}_2}^2} -
   t^*_1\,{{\bar{w}_1}^2}\,{{\bar{w}_2}^2} -
   2\,t^*_2\,{{\bar{w}_1}^3}\,{{\bar{w}_2}^2} +
   t^*_1\,\bar{w}_0\,{{\bar{w}_2}^3} = 0,
\end{array}
\end{equation}
\begin{equation}\label{ex8eq2}\everymath{\displaystyle}
\begin{array}{l}
-2\,t_2\,\bar{w}_0\,{{\bar{w}_1}^2} +
   2\,t_2\,{{\bar{w}_0}^2}\,\bar{w}_2 +
   t_1\,\bar{w}_0\,\bar{w}_1\,\bar{w}_2 +
   4\,t_2\,\bar{w}_0\,w_1\,\bar{w}_1\,
    \bar{w}_2 - t_0\,\bar{w}_0\,{{\bar{w}_2}^2}\\  \\
     -   t^*_1\,\bar{w}_0\,\bar{w}_1\,{{\bar{w}_2}^2} -
   2\,t^*_2\,\bar{w}_0\,{{\bar{w}_1}^2}\,
    {{\bar{w}_2}^2} - 2\,t^*_2\,{{\bar{w}_0}^2}\,
    {{\bar{w}_2}^3} = 0.
\end{array}
\end{equation}
Identification of  powers of $p$ in \eqref{hod1} implies
\begin{equation}\label{ex8eq3eq4}\everymath{\displaystyle}\begin{array}{rcl}
p: &  &
-2\,t_2 + \bar{w}_2 = 0,
\\  \\
p^0: &  &
2\,t_2\,\bar{w}_1 - t_1\,\bar{w}_2 -
   4\,t_2\,w_1\,\bar{w}_2 +
   w_1\,{{\bar{w}_2}^2} = 0,
  \\  \\
p^{-1} : &  &
-2\,t_2\,{{\bar{w}_1}^2} +
   2\,t_2\,\bar{w}_0\,\bar{w}_2 +
   t_1\,\bar{w}_1\,\bar{w}_2 +
   4\,t_2\,w_1\,\bar{w}_1\,\bar{w}_2 -
   t_0\,{{\bar{w}_2}^2} - t^*_1\,\bar{w}_1\,
    {{\bar{w}_2}^2} \\  \\
    &   &- 2\,t^*_2\,{{\bar{w}_1}^2}\,
    {{\bar{w}_2}^2} - 4\,t^*_2\,\bar{w}_0\,
    {{\bar{w}_2}^3} + w_2\,{{\bar{w}_2}^3} =
    0.
\end{array}\end{equation}
By solving equations \eqref{ex8eq1}-\eqref{ex8eq3eq4}
we get the solution:
$$\bL=
p + {\frac{t^*_1 + 2\,t_1\,t^*_2}
    {1 - 4\,t_2\,t^*_2}} + {\frac{2\,t_0\,t^*_2}
    {p\,\left( 1 - 4\,t_2\,t^*_2 \right) }},$$
$$\frac{1}{\bar{\bL}}=
 \frac{t_0}{p\,\left( 1 - 4\,t_2\,t^*_2 \right) } +
  \frac{t_1+ 2\,t^*_1\,t_2}{1 - 4\,t_2\,t^*_2}+2\,p\,t_2 ,$$
which leads to the conformal map describing  an \emph{ellipse
growing from a circle} \cite{6}
\begin{equation}\label{el}
z=\Big(\frac{t_0}{ 1 - 4\,t_2\,t^*_2 }\Big)^{\frac{1}{2}}\,p+
\frac{t_1^*+2t_1\,t_2^*}{ 1 - 4\,t_2\,t^*_2 }+2\,
\Big(\frac{t_0}{ 1 - 4\,t_2\,t^*_2 }\Big)^{\frac{1}{2}}\,
\frac{t_2^*}{p}.
\end{equation}

\subsection*{II) $n_1=n_2=3$ (the hypotrochoid)}

Let us take $t_1=t_2=t^*_1=t^*_2=0$, then
$p\,\bar{\bL}^{-1}$ is a third degree polynomial and \eqref{hod2}
gives rise to a system of three equations  which can be reduced
to
\begin{equation}\label{ex9eq1}\everymath{\displaystyle}
\begin{array}{l}
3\,t_3\,{{\bar{w}_2}^4} -
   9\,t_3\,\bar{w}_1\,{{\bar{w}_2}^2}\,
    \bar{w}_3 - 9\,t_3\,w_1\,
    {{\bar{w}_2}^3}\,\bar{w}_3 +
   3\,t_3\,{{\bar{w}_1}^2}\,{{\bar{w}_3}^2} +
   6\,t_3\,\bar{w}_0\,\bar{w}_2\,
    {{\bar{w}_3}^2}\\  \\
     + 18\,t_3\,w_1\,
    \bar{w}_1\,\bar{w}_2\,{{\bar{w}_3}^2} +
   9\,t_3\,{{w_1}^2}\,{{\bar{w}_2}^2}\,
    {{\bar{w}_3}^2} + 9\,t_3\,w_2\,
    {{\bar{w}_2}^2}\,{{\bar{w}_3}^2} -
   9\,t_3\,\bar{w}_0\,w_1\,{{\bar{w}_3}^3}\\  \\
    -   9\,t_3\,{{w_1}^2}\,\bar{w}_1\,
    {{\bar{w}_3}^3} - 9\,t_3\,\bar{w}_1\,
    w_2\,{{\bar{w}_3}^3} -
   t_0\,\bar{w}_2\,{{\bar{w}_3}^3} -
   3\,t^*_3\,{{\bar{w}_1}^3}\,\bar{w}_2\,
    {{\bar{w}_3}^3}\\  \\
     - 18\,t^*_3\,\bar{w}_0\,
    \bar{w}_1\,{{\bar{w}_2}^2}\,{{\bar{w}_3}^3} +
   9\,t^*_3\,\bar{w}_0\,{{\bar{w}_1}^2}\,
    {{\bar{w}_3}^4} = 0
\end{array}
\end{equation}
\begin{equation}\label{ex9eq2}\everymath{\displaystyle}
\begin{array}{l}
3\,t_3\,\bar{w}_1\,{{\bar{w}_2}^3} -
   6\,t_3\,{{\bar{w}_1}^2}\,\bar{w}_2\,
    \bar{w}_3 - 3\,t_3\,\bar{w}_0\,
    {{\bar{w}_2}^2}\,\bar{w}_3 -
   9\,t_3\,w_1\,\bar{w}_1\,{{\bar{w}_2}^2}
\,
    \bar{w}_3 + 6\,t_3\,\bar{w}_0\,\bar{w}_1\,
    {{\bar{w}_3}^2} \\  \\
    + 9\,t_3\,w_1\,
    {{\bar{w}_1}^2}\,{{\bar{w}_3}^2} +
   9\,t_3\,\bar{w}_0\,w_1\,\bar{w}_2\,
    {{\bar{w}_3}^2} + 9\,t_3\,{{w_1}^2}\,
    \bar{w}_1\,\bar{w}_2\,{{\bar{w}_3}^2} +
   9\,t_3\,\bar{w}_1\,w_2\,\bar{w}_2\,
    {{\bar{w}_3}^2}\\  \\
     - 9\,t_3\,\bar{w}_0\,
    {{w_1}^2}\,{{\bar{w}_3}^3} -
   t_0\,\bar{w}_1\,{{\bar{w}_3}^3} -
   3\,t^*_3\,{{\bar{w}_1}^4}\,{{\bar{w}_3}^3} -
   9\,t_3\,\bar{w}_0\,w_2\,{{\bar{w}_3}^3} -
   18\,t^*_3\,\bar{w}_0\,{{\bar{w}_1}^2}\,
    \bar{w}_2\,{{\bar{w}_3}^3} = 0
\end{array}
\end{equation}
\begin{equation}\label{ex9eq3}\everymath{\displaystyle}
\begin{array}{l}
3\,t_3\,\bar{w}_0\,{{\bar{w}_2}^3} -
   6\,t_3\,\bar{w}_0\,\bar{w}_1\,\bar{w}_2\,
    \bar{w}_3 - 9\,t_3\,\bar{w}_0\,w_1\,
    {{\bar{w}_2}^2}\,\bar{w}_3 +
   3\,t_3\,{{\bar{w}_0}^2}\,{{\bar{w}_3}^2}\\  \\
    +
   9\,t_3\,\bar{w}_0\,w_1\,\bar{w}_1\,
    {{\bar{w}_3}^2} + 9\,t_3\,\bar{w}_0\,
    {{w_1}^2}\,\bar{w}_2\,{{\bar{w}_3}^2} +
   9\,t_3\,\bar{w}_0\,w_2\,\bar{w}_2\,
    {{\bar{w}_3}^2} - t_0\,\bar{w}_0\,{{\bar{w}_3}^3}\\  \\
     -   3\,t^*_3\,\bar{w}_0\,{{\bar{w}_1}^3}\,
    {{\bar{w}_3}^3} - 18\,t^*_3\,{{\bar{w}_0}^2}\,
    \bar{w}_1\,\bar{w}_2\,{{\bar{w}_3}^3} -
   6\,t^*_3\,{{\bar{w}_0}^3}\,{{\bar{w}_3}^4} = 0
\end{array}
\end{equation}
Now, by equating coefficients of  powers of $p$ in \eqref{hod1} we obtain
\begin{equation}\label{ex9eq4eq5}\everymath{\displaystyle}\begin{array}{rcl}
p: &  &
-3\,t_3 + \bar{w}_3 = 0
\\  \\
p^0: &  &
3\,t_3\,\bar{w}_2 -
   9\,t_3\,w_1\,\bar{w}_3 +
   w_1\,{{\bar{w}_3}^2} = 0
  \\  \\
p^{-1} : &  &
-3\,t_3\,{{\bar{w}_2}^2} +
   3\,t_3\,\bar{w}_1\,\bar{w}_3 +
   9\,t_3\,w_1\,\bar{w}_2\,\bar{w}_3 -
   9\,t_3\,{{w_1}^2}\,{{\bar{w}_3}^2}\\  \\
   &   & -   9\,t_3\,w_2\,{{\bar{w}_3}^2} +
   w_2\,{{\bar{w}_3}^3} = 0
  \\  \\
p^{-2} : &  &
3\,t_3\,{{\bar{w}_2}^3} -
   6\,t_3\,\bar{w}_1\,\bar{w}_2\,\bar{w}_3 -
   9\,t_3\,w_1\,{{\bar{w}_2}^2}\,\bar{w}_3 +
   3\,t_3\,\bar{w}_0\,{{\bar{w}_3}^2} \\  \\
  &   & +   9\,t_3\,w_1\,\bar{w}_1\,{{\bar{w}_3}^2} +
   9\,t_3\,{{w_1}^2}\,\bar{w}_2\,
    {{\bar{w}_3}^2} + 9\,t_3\,w_2\,
    \bar{w}_2\,{{\bar{w}_3}^2} - t\,{{\bar{w}_3}^3}\\  \\
    &   & -   3\,t^*_3\,{{\bar{w}_1}^3}\,{{\bar{w}_3}^3} -
   18\,t^*_3\,\bar{w}_0\,\bar{w}_1\,\bar{w}_2\,
    {{\bar{w}_3}^3} - 9\,t^*_3\,{{\bar{w}_0}^2}\,
    {{\bar{w}_3}^4} + w_3\,{{\bar{w}_3}^4} = 0
\end{array}\end{equation}
Now by setting $w_1=\bar{w}_1=w_2=\bar{w}_2=0$ one finds the
solution
$$\bL=
p + {\frac{3\,t^*_3\,{{\bar{w}_0}^2}}{{p^2}}},\quad
\frac{1}{\bar{\bL}}=3\,{p^2}\,t_3 + {\frac{\bar{w}_0}{p}},$$
with
$$\bar{w}_0={\frac{1 - {\sqrt{1 - 72\,t_0\,t_3\,t^*_3}}}
   {36\,t_3\,t^*_3}},$$
which satisfies (58) and leads to the conformal map associated
with the hypotrochoid \cite{23}
\begin{equation}\label{hi}
z=\bar{w}_0^{\frac{1}{2}}\,p+\frac{3\,t^*_3\,\bar{w}_0}{p^2}.
\end{equation}

\noindent {\bf Acknowledgements}

The authors are grateful to Prof. Manuel Ma\~{n}as for many useful
discussions.

\end{document}